\documentclass[fleqn,usenatbib]{mnras}

\usepackage{newtxtext,newtxmath}

\usepackage[T1]{fontenc}

\DeclareRobustCommand{\VAN}[3]{#2}
\let\VANthebibliography\thebibliography
\def\thebibliography{\DeclareRobustCommand{\VAN}[3]{##3}\VANthebibliography}

\usepackage{graphicx}	
\usepackage{amsmath}	
\usepackage{xcolor}

\newcommand{\hd}[1]{}

\newcommand{\tys}[1]{}
\newcommand{\ty}{}

\newcommand{\Sighi}{\Sigma_{\mathrm{HI}}}
\newcommand{\HI}{\ensuremath{\mathrm{H}\scriptstyle\mathrm{I}}}
\newcommand{\sighi}{\sigma_{\mathrm{HI}}}

\newcommand{\vcirc}{v_{\mathrm{circ}}}
\newcommand{\vproj}{v_{\mathrm{proj}}}

\newcommand{\mbar}{M_{\mathrm{bar}}}
\newcommand{\msun}{M_{\odot}}

\newcommand{\mhi}{M_{\mathrm{HI}}}
\newcommand{\rb}{\overline{r}_b}
\newcommand{\mtot}{M_{\text{tot}}}
\newcommand{\dkl}{D_{\text{KL}}}

\newcommand{\zratio}{\mathcal{Z}_{\textrm{asym}}/\mathcal{Z}_{\textrm{sym}}}

\title[Halo properties from integrated \HI{} profiles]{Dark matter halo properties from spatially integrated \HI{} flux profiles}

\author[Yasin \& Desmond]
{Tariq Yasin$^{1}$\thanks{\href{mailto:tariq.yasin@physics.ox.ac.uk}{tariq.yasin@physics.ox.ac.uk}}
and Harry Desmond$^{2}$
\\
$^1$Astrophysics, University of Oxford, Denys Wilkinson Building, Keble Road, Oxford, OX1 3RH, UK\\
$^2$Institute of Cosmology \& Gravitation, University of Portsmouth, Dennis Sciama Building, Portsmouth, PO1 3FX, UK\\
}

\date{Accepted XXX. Received YYY; in original form ZZZ}

\pubyear{2015}

\begin{document}
\label{firstpage}
\pagerange{\pageref{firstpage}--\pageref{lastpage}}
\maketitle

\begin{abstract}
Resolved rotation curves (RCs) are our best probe of the dark matter distribution around individual galaxies. However their acquisition is resource-intensive, rendering them impractical for large-scale surveys and studies at higher redshift. Spatially integrated \HI{} flux profiles on the other hand are observationally abundant and also probe dynamics across the whole \HI{} disc. Despite this, they are typically only studied using the highly compressed linewidth summary statistic, discarding much of the available information. Here we construct a Bayesian
model to infer halo properties from the full shape of the spatially integrated 21-cm line profile of a galaxy, utilising all the available information.
We validate our model by assessing the consistency of halo parameters obtained from the flux profile with those obtained from RC fits for galaxies where both are available,
finding good agreement provided the profile is not
strongly asymmetric.
We study the relative constraining power (quantified using the Kullback--Leibler divergence of the posterior from the prior), finding the flux profile inference recovers posteriors on generalised Navarro--Frenk--White halo parameters on average three times tighter than those from the linewidth, and in some cases as tight as those from resolved RCs.
Finally we introduce and validate a probabilistic empirical model for the spatial distribution of \HI{}, enabling our model to be applied to datasets for which no spatially resolved \HI{} information is available.
As the next-generation of \HI{} observatories comes online, our framework will enable mass modelling in new regimes,
with particular utility for constraining the dark matter content of galaxies across cosmic time.
\end{abstract}




\section{Introduction}
In the standard model of cosmology, galaxies form from the condensation of baryons in the centre of dark matter haloes. In heavy enough halos, the baryon density at the centre of the halo becomes high enough to form stars and eventually the luminous galaxies we observe today \citep{whiteCoreCondensationHeavy1978}. Galaxies then grow through either the continuous formation of stars from (\emph{in situ}) and through the mergers of their dark matter halos (\emph{ex situ}) under hierarchical structure formation. This suggests that the internal properties of galaxies and their spatial distribution should be closely related to the properties and distributions of the underlying dark matter halos. The relationship between galaxies and their halos has been termed the \emph{galaxy--halo} connection \citep[reviewed in][]{wechslerConnectionGalaxiesTheir2018}, and constraining it is a key problem in modern astrophysics.

There are numerous ways to constrain dark matter halo properties and the galaxy--halo connection more generally. These include empirical methods such as abundance matching \citep{wechslerClusteringHighredshiftGalaxies1999} and halo occupation distribution modelling \citep{coorayHaloModelsLarge2002}, and dynamical methods using the kinematics of stars and gas \citep{sofueRotationCurvesSpiral2001} as well as weak \citep{mandelbaumGalaxyHaloMasses2014} and strong \citep{limousinStrongLensingAbell2008} lensing. Of these, the dynamics of neutral hydrogen \HI{}, found predominantly in late-type galaxies, is particularly powerful: it forms a dynamically cold rotating disc, resulting in a relatively simple connection between the observed velocity of the gas and the underlying circular velocity tracing the gravitational potential. \HI{} discs also extend far out beyond the optical radius of the galaxy and into the region where dark matter is most apparent. This is particularly important for constraining global properties of halos, which are much larger than galaxies.

Spatially resolved spectroscopic studies of \HI{} produce rotation curves (RCs), the rotation speed of the gas as a function of radius from the galaxy centre. This enables relatively tight constraints on dark matter halo properties demanding both high resolution and sensitivity. The largest collations of \HI{} RC observations today number in the 100s \citep{lelliSPARCMASSMODELS2016}, which is set to expand to thousands over the coming years with the arrival of instruments such as the Square Kilometer Array \citep{koribalskiWALLABYSKAPathfinder2020}.

Dark matter halo properties are typically constrained in dynamical studies by assuming a parameterised halo density profile, which is combined with \emph{mass models} describing the mass distribution of baryons to forward model the gravitational potential. The parameters of the halo (and auxiliary galaxy parameters such as mass-to-light ratios, distance and inclination) can then be constrained by observational data. Foremost among the density profiles is the the Navarro--Frenk--White profile (NFW, \citealt{navarroUniversalDensityProfile1997}), which is an approximately universal two-parameter function describing halos in cold dark matter-only simulations of structure formation. However, observations of dwarf galaxies have found some galaxies are better described by a central core in the dark matter density \citep{deblokHIObservationsLow1996}, inspiring an array of different halo profiles inspired by baryonic interactions with dark matter (which may flatten cusps, e.g.~\citealt{burkertStructureDarkMatter1995}) or exotic dark matter models \citep{yangParametricModelSelfinteracting2024}. Typically halo profiles are described by a total halo mass and a concentration parameter (describing how spatially concentrated the dark matter is), as well as potentially additional parameters governing the profile shape. 

Whilst rotation curves can give strong constraints on halo parameters, they suffer from a number of limitations. Observations need both sufficient spatial resolution to resolve the \HI{} disc and long integration times to achieve adequate signal-to-noise in each radial bin, rendering them extremely expensive to obtain for large samples of galaxies, small and faint objects, and high redshift objects (due to cosmological surface brightness dimming).  Although upcoming surveys such as Apertif-Medium deep on the Westerbrook Synthesis Radio Telescope \citep{vancappellenApertifPhasedArray2022}, MIGHTEE on MeerKat \citep{maddoxMIGHTEEHIHIEmission2021} and the Square Kilometre Array (SKA) will provide dramatic improvements in instrumental resolution and sensitivity, deriving RCs will remain challenging or impractical in many regimes of scientific interest.
For example, \HI{} RCs will still only be accessible out to z$\sim$0.5 \citep{blythExploringNeutralHydrogen2015}, even with the unprecedented capabilities of the SKA. Similarly, while it is possible to resolve the RCs of the most nearby low-mass dwarf galaxies, this comes at great observational cost, requiring deep interferometric follow up on candidate galaxies with the VLA, Meerkat or the SKA, severely limiting sample sizes \citep{mancerapinaNoNeedDark2022, reyEDGEDarkMatter2023}. These limitations prevent us from leveraging RCs to statistically probe critical aspects of galaxy formation, especially its faint end and redshift evolution.  

An alternate probe of \HI{} dynamics is spatially integrated 21-cm flux profiles, which are observationally cheap and available for large samples of galaxies \citep{meyerHIPASSCatalogueData2004,haynesAreciboLegacyFast2018, zhangFASTAllSky2024}. Typically flux profiles are compressed to the linewidth summary statistic, defined as the width of the profile at some fraction of its mean or peak flux. Linewidths have been shown to give interesting constraints on the halo properties of galaxies \citep{leismanAlmostDarkGalaxies2017, yasinInformationHaloProperties2023, yasinInferringDarkMatter2023}, and are also widely used in dynamical tests such as Tully--Fisher studies \citep{lelliBaryonicTullyFisher2019,ponomarevaMIGHTEEHBaryonicTully2021,brooksNorthsouthAsymmetryALFALFA2023,sardoneClosingGapObserved2023}. As highlighted in \citet{yasinInformationHaloProperties2023}, whilst the linewidth can be relatively constraining at low mass where the dark matter halo is dominant at all radii, for high mass baryon-dominated galaxies there is a degeneracy in the linewidth between galaxies with and without flat RCs. For these galaxies the peak circular velocity is due predominantly to the mass of the stellar disc, and so the flux profile is similarly wide irrespective of whether the RC then declines in the outskirts. Particular in this regime, analyses that simply use the linewidth are prone to lose a lot of information. This is especially pertinent when attempting to probe out to higher redshift, as it is the massive late-type spirals that are seen to the greatest distances.

Ideally the full information content of the \HI{} flux profile would be used to constrain halo properties, without compression into a linewidth. The shape of the full flux profile has previously been highlighted as a potentially powerful probe of the dynamics through studies based on semi-analytical models and cosmological simulations \citep{2021arXiv210504570P,el-badryGasKinematicsFIRE2018a}. In \citet{yasinConsistencyRotationCurves2024}, we took the first steps towards using the information of the full profile to constrain galaxy dynamics by demonstrating good
consistency between RCs and integrated flux profiles under basic modelling assumptions.

In this contribution we study the power of flux profiles to constrain the dark matter distribution around galaxies by fitting dark matter halo models, comparing and contrasting the constraining power with that of RCs. In particular, we ask
\begin{enumerate}
    \item Are the constraints obtained by integrated flux profiles and RCs on halo properties consistent?
    \item What is the relative information content on halo parameters of integrated \HI{} flux profiles compared to RCs?
    \end{enumerate}

In Section \ref{sec:data} we describe the observational data used. In Section \ref{sec:methods} we describe our method. In Section \ref{sec:results} we present our constrains on halo properties from integrated flux profiles. In Section \ref{sec:discussion} we discuss our findings, and we conclude in Section \ref{sec:conclusions}. Throughout the paper we assume $H_0 = 70$ km/s/Mpc.


\section{Data}\label{sec:data}

\subsection{SPARC}\label{sec:SPARC}

We utilise the SPARC database\footnote{\url{http://astroweb.cwru.edu/SPARC/}}, which offers \(\text{H\textsc{i}}\) rotational curves for a selection of 175 late-type galaxies \citep{lelliSPARCMASSMODELS2016}. We also use the azimuthally-averaged \(\text{H\textsc{i}}\) surface density profiles of the SPARC galaxies (F. Lelli, priv. comm.). The database is a compilation of rotation curves gathered over multiple decades, and primarily consists of \HI{} data, supplemented by H\(\alpha\) observations for the inner regions of some galaxies. The SPARC galaxies are diverse, featuring luminosities from \(10^7 L_{\odot}\) to \(10^{12} L_{\odot}\), surface brightness values between about 5 and \(5000 L_{\odot} \, \mathrm{pc}^{-2}\), and \(\text{H\textsc{i}}\) masses ranging from \(10^7\) to \(10^{10.6} M_{\odot}\). They also display a diverse array of galactic morphologies. We use only galaxies with quality flag 1 (\emph{high}) and 2 (\emph{medium}).

\subsection{ALFALFA}\label{sec:ALFALFA}

Integrated flux measurements were taken from the ALFALFA\footnote{\url{http://egg.astro.cornell.edu/alfalfa/data/index.php}} \citep{haynesAreciboLegacyFast2018}, a blind survey with the Arecibo telescope across approximately 7000 \(\text{deg}^2\) of the Northern sky. ALFALFA is able to detect sources up to a redshift of $\sim0.06$. The complete \(\alpha.100\) catalogue lists around 31,500 extragalactic objects, spanning \(\text{H\textsc{i}}\) masses from \(10^6\) to \(10^{11}\) \(\textrm{M}_{\odot}\). For each object, the provided data includes a spectrum with the baseline removed, segmented into velocity bins of approximately 5 km/s.

Cross-referencing was conducted with optical positions from SPARC, sourced from the NASA/IPAC Extragalactic Database\footnote{\url{https://ned.ipac.caltech.edu}}, using a 4-arcsecond matching criterion. This crossmatch revealed 40 galaxies common to both datasets, which were checked by comparing \(\text{H\textsc{i}}\) masses and linewidths between SPARC and ALFALFA. During this process, we excluded NGC4214 as the \HI{} mass reported by the SPARC and THINGS surveys is approximately 10 times greater than the ALFALFA value (assuming consistent distance measurements), and there is also milder disagreement with the ALFALFA linewidth.

The standard ALFALFA pipeline is designed to accurately recover flux profiles for unresolved sources. However, many SPARC galaxies are relatively nearby, resulting in angular sizes significantly larger than the ALFALFA beam width. This can lead to an underestimation of flux from the outer regions of these galaxies, potentially causing one or both horns of the flux profile to appear artificially diminished.

To address this issue, \citet{hoffmanTotalALFALFANeutral2019} developed an alternative ALFALFA analysis pipeline to enable precise flux extraction for galaxies spanning more than a few arcminutes on the sky. Through visual inspection, we confirmed that this reanalysis successfully captured more flux from the galactic outskirts, resulting in more prominent horns for many of the reexamined galaxies. However, the complete spectrum data was not retained for all galaxies reanalysed using this alternative method, which primarily focused on total flux measurements. Consequently, our final sample comprises 20 galaxies, evenly split between two categories: 10 galaxies for which the reanalysed spectra were available (L. Hoffman, priv. comm.), and 10 galaxies with angular extents small enough that the reanalysis pipeline was unnecessary. This is the same sample as used in \citet{yasinConsistencyRotationCurves2024}.

\section{Methods}\label{sec:methods}

\subsection{Dynamical modelling}

\subsubsection{Rotation curve model}

We wish to constrain dark matter halo properties using both SPARC rotation curves and ALFALFA flux profiles. Both of these first require that we forward model the projected rotational velocity as a function of radius $\vproj(r)$ for each galaxy.

This is done by combining a parameterised halo profile with mass models describing the distribution of baryons in the galaxy to forward model the circular velocity profile. The total circular velocity is

\begin{equation}\protect\hypertarget{eq:ph}{}{
V_{\mathrm{c}}^{2}(r)=V_{\mathrm{DM}}^2 + \Upsilon_{\text{bulge}}V_{\mathrm{bulge}}^2 +  \Upsilon_{\text{disc}}V_{\mathrm{disc}}^2
 + V_{\mathrm{gas}}|V_{\mathrm{gas}}|,
}\label{eq:ph}
\end{equation}
where $V_{\mathrm{bulge}},V_{\mathrm{disc}}$ and $V_{\mathrm{gas}}$ are the circular velocity contributions of the bulge, disc and gas respectively (the latter can be negative due to holes in the centre of the gas distribution). These are calculated from photometry and are tabulated in the SPARC database. The mass-to-light ratios $\Upsilon_{\text{bulge}}$ and $\Upsilon_{\text{disc}}$ are free parameters in the inference (although only a fraction of galaxies have a bulge), with lognormal priors based on the fiducial SPARC model of $\mu_{\text{bulge}}=0.5$ and $\mu_{\text{disc}}=0.7$, each with 0.1 dex scatter.

For the dark matter we assume a generalised NFW (gNFW) profile \citep{wyitheGravitationalLensStatistics2001}

\begin{equation}
\rho_{\mathrm{gNFW}}(r)=\frac{\rho_{s}}{\left(\frac{r}{r_{s}}\right)^{\alpha}\left[1+\left(\frac{r}{r_{s}}\right)\right]^{3-\alpha}}.
\label{eq:gnfw}\end{equation}
This reduces to NFW for $\alpha=1$. For lower values of $\alpha$ the inner region becomes \emph{cored}, which is found to be a better match to many SPARC galaxies \citep{liComprehensiveCatalogDark2020}. The mass enclosed is given by
\begin{equation}
M_{\mathrm{DM}}(r) =4 \pi \rho_s r_{s}^{3}\left[ B(x/(1+x),3-\alpha,0) \right],
\end{equation}
where $B(z ; a, b) \equiv \int_0^z u^{a-1}(1-u)^{b-1} d u$. This gives a circular velocity contribution of

\begin{equation}\protect{}{
\frac{V_{\text{DM}}(r)}{V_{\text{halo}}} = \sqrt{\frac{M_{\text{DM}}(r)}{M_{\text{halo}}} \frac{R_{\text{halo}}}{r}},
}\end{equation}
where $M_{\text{halo}}$ is the total halo mass, $R_{\text{halo}}$ is the virial radius and $V_{\text{halo}}$ is the circular velocity at the virial radius.

We then assume that the rotational velocity is equal to the circular velocity (i.e. that we ignore pressure support of the gas), which is expected to be a good assumption for these galaxies, which have gas dispersions $\sighi$ of around 10 km/s \citep{mogotsiHICOVelocity2016}, much lower than typical rotation velocities. Then $\vcirc$ is projected along the line of sight by the inclination to get the projected rotation velocity
\begin{equation}
    \vproj=\vcirc \sin i,
\end{equation}
where $i$ is inclination ($i=0^\circ$ edge on; $i=90^\circ$ face on).

The circular velocity contribution of the baryonic components and the observed radius scales with the assumed distance (which is a free parameter in the fit) as
\begin{equation}
V_{\text{disc,bulge,gas}}(r) \propto \sqrt{D},
\end{equation}
and the physical radius scales with the assumed $D$
\begin{equation}
r \propto D.
\end{equation}
When using the observed rotation curve to constrain halo properties, $\vproj$ can be compared to the data to infer halo and nuisance parameters describing the galaxy with the likelihood
\begin{equation}\label{eq:likelihood}\protect{}{
\mathcal{L_{\text{V}}}(D|\theta,\mathcal{M}) = {\displaystyle \prod_{i}} \frac{\exp\{
-(V_{\text{i,obs}} - v_{\text{proj}}(r_\text{i}))^2 / (2(\delta V_{\text{i,obs}}^2 + \sigma_V^2) \}}
{\sqrt{2\pi\delta (V_{\text{i,obs}}^2 + \sigma_V^2)}} },
\end{equation}
where $V_{\text{i,obs}}$ is the $i$th observed velocity, $\delta V_{\text{i,obs}}$ its observational uncertainty and $\sigma_V$ is an additional uncorrelated Gaussian noise term which is a free parameter in our model. This is necessary because some SPARC RC uncertainties are underestimated \citep{liComprehensiveCatalogDark2020,sellwoodUncertaintiesGalaxyRotation2021,zentnerCriticalAssessmentSolutions2022}, leading to extremely tight constraints (despite poor model fits) that are likely biased. The introduction of $\sigma_V$ effectively broadens the uncertainties to allow the model to fit the data well, helping to make the analysis conservative.

\subsubsection{Integrated 21-cm model}\label{sec:integrated_model}

To use the \HI{} flux profile to constrain halo profile parameters instead of the RC, the model $\vproj$ is combined with a surface density as a function of radius $\Sighi$ under the assumption of axisymmetry to calculate a model flux profile $\Psi_{\mathrm{pred}}$. We utilise the method of \cite{obreschkowSIMULATIONCOSMICEVOLUTION2009}, modelling the \HI{} gas disc as a set of concentric, infinitely thin rings, each rotating with a circular velocity $v_c(r)$ and having a surface density $\Sigma_{\mathrm{HI}}(r)$. The normalised flux emitted at a specific velocity from a single ring is

\[
\tilde{\psi}\left(v_{\lambda}, v_{\mathrm{proj}}\right) = \left\{
\begin{array}{ll}
\frac{1}{\pi \sqrt{v_{\mathrm{proj}}^{2} - v_{\lambda}^{2}}} & \text{if } |v_{\lambda}| < v_{\mathrm{proj}} \\
0 & \text{otherwise}
\end{array}
\right.
\]
where $v_{\lambda}$ represents the velocity relative to the galaxy's barycentre, corresponding to an observed wavelength $\lambda$, and the normalisation ensures that $\int \tilde{\psi} dv = 1$. To account for the random motion of the gas, we introduce a velocity dispersion $\sigma_{\mathrm{HI}}$, which also smooths out the singularity that occurs when $v_{\mathrm{proj}} = v_{\lambda}$. The resulting normalised flux from each ring is

\[
\psi\left(v_{\lambda}, v_{\mathrm{proj}}\right) = \frac{1}{\sigma_{\mathrm{HI}} \sqrt{2\pi}} \int_{-\infty}^{\infty} dV \exp\left[-\frac{(v_{\lambda} - V)^{2}}{2\sigma_{\mathrm{HI}}^{2}}\right] \tilde{\psi}\left(V, v_{\mathrm{proj}}\right)
\]
where $\sigma_{\mathrm{HI}}$ is the Gaussian dispersion width. The total flux per unit velocity for the entire galaxy is obtained by integrating $\psi$ over the radial extent of the galaxy, weighted by the surface density
\[
\psi_{\mathrm{tot}}\left(v_{\lambda}\right) = \frac{2\pi}{M_{\mathrm{HI}}} \int_{0}^{\infty} r \Sigma_{\mathrm{HI}}(r) \psi\left(v_{\lambda}, v_{\mathrm{proj}}(r)\right) dr,
\]
with normalisation ensuring $\int \psi_{\mathrm{tot}} dv = 1$. We convolve the profile with a Gaussian kernel with 5 km/s full-width-half maximum equal corresponding to the instrumental resolution. We calculate observed 21-cm flux using the standard conversion $\Psi_{\mathrm{tot}} = \frac{M_{\mathrm{HI}}}{D^2}  \frac{\mathrm{ Jy \,  kms^{-1} \, Mpc^2 M_{\odot}^{-1}}}{2.356 \times 10^5}$ \citep{haynesAreciboLegacyFast2018}, where $\Psi_{\mathrm{tot}}$ is in Jy km/s, $D$ is the assumed distance in Mpc and $M_{\mathrm{HI}}$ is in $M_{\odot}$). \HI{} observations typically have a calibration uncertainty of 10\% in flux \citep{haynesAreciboLegacyFast2018}, which we account for by multiplying the predicted flux by a new parameter $C$, which we give a Gaussian prior with mean 1 and width 0.1. The likelihood is then
\begin{equation}\label{eq:hi_likelihood}\protect{}{
\mathcal{L}_{\Psi}(\mathcal{D}|\theta) = {\displaystyle \prod_{j}^{n}} \frac{\exp\{
-(\Psi_{\text{j,obs}} - C \Psi_{\text{pred}}(v_{\lambda,j},\vproj))^2 / (2 \sigma_{\mathrm{tot}}^2) \}}
{\sqrt{2\pi}\sigma_{\mathrm{tot}}} },
\end{equation}
where $\Psi_{\text{j,obs}}$ is the observed flux in bin $j$, $\Psi_{\text{pred}}$ is the predicted flux assuming perfect calibration, and $\sigma_{\mathrm{tot}} = \sqrt{\sigma_\text{rms}^2 + \sigma_{\Psi}^2}$ with $\sigma_\text{rms}$ the spectrum RMS noise and $\sigma_{\Psi}$ an additional noise term analogous to $\sigma_V$, which we again leave as a free parameter in the inference. As $C$ is perfectly degenerate with $\mhi$ we can analytically marginalise over it to give
\begin{equation}\label{eq:hi_likelihood}\protect{}{
\mathcal{L}_{\Psi}(\mathcal{D}|\theta) = {\displaystyle \prod_{j}^{n}} \frac{\exp\{
-(\Psi_{\text{j,obs}} - \Psi_{\text{pred}}(v_{\lambda,j},\vproj))^2 / (2 \sigma_m^2) \}}
{\sqrt{2\pi}\sigma_m} },
\end{equation}
with
\begin{equation}
\sigma_m \equiv (0.1 \Psi_{\text{pred}})^2  + \sigma_\text{rms}^2 + \sigma_{\Psi}^2.
\end{equation}

\subsubsection{Fitting procedure}\label{sec:inference}

Bayes theorem gives the posterior probability on model parameters given some set of data $\mathcal{D}$ and the prior $\pi$

\begin{equation}
\mathcal{P}(\theta |\mathcal{D})=\frac{\mathcal{L}(\mathcal{D}|\theta) \pi(\theta)}{\mathcal{Z}}.
\end{equation}
The normalisation factor $\mathcal{Z}$ is the Bayesian evidence
\begin{equation}
    \mathcal{Z} = \int \mathcal{L}(\mathcal{D}|\theta) \pi(\theta) \operatorname{d}\theta,
\end{equation}
which is the probability for the data $\mathcal{D}$ under a chosen model. The \emph{Bayes factor} $\mathcal{Z}_A/\mathcal{Z}_B$, where $A$ and $B$ are two separate models constrained using the same dataset, is the relative probability of the two models and hence compares them on the data. We use the \texttt{Multinest} algorithm \citep{ferozMultiNestEfficientRobust2009,ferozImportanceNestedSampling2019} to sample model parameters and calculate the evidence with 500 live points, a sampling efficiency of 0.3, and an evidence tolerance of 0.5. We find our results are robust to increasing the live points, and reducing the evidence tolerance \ty{to do}. \ty{also prior dependence}

We summarise the model parameters and their priors in Table \ref{tab:parameters}. The halo parameters ($\mtot$, $c$, $\alpha$) and galaxy parameters ($\Upsilon_{\mathrm{disc/bulge}}$, $D$ and $i$) are present in all models. $\mhi$ and $\sigma_{HI}$ are only included when constraining with the flux profile. In particular $\mhi$ dictates the normalisation of the flux profile. Throughout this work we fix $V_{\mathrm{gas}}$ to the values tabulated in SPARC, independent of the $\mhi$ inferred from the flux profile.

In Section \ref{sec:probgas} we introduce a new \emph{probabilistic} model for $\Sighi$ (as an alternative to assuming the SPARC $\Sighi$). This introduces two new parameters  $\Sigma_0$ and $\rb$, which are only included for inferences where the probabilistic model is used. The additional RC (flux profile) noise term $\sigma_V$ ($\sigma_{\Psi}$) is only included when the inference is constrained with the RC (flux profile) data.

\begin{table*}
\caption{\label{tab:parameters}The free parameters needed for the rotation curve and flux profile fits, their physical definitions and their Bayesian priors. The uniform priors on non-halo properties are chosen to be uninformative. The first 7 parameters are used in all fits. $\mhi$ and $\sighi$ are only included for flux profile fits. Additionally $\Sigma_0$, $\overline{r}_b$ are included when using the probabilistic model for $\Sighi$ to calculate the observed flux (Section \ref{sec:probgas}). The additional noise parameters are included in all fits with their corresponding data. Except for the halo parameters, all uniform priors are uninformative. $\mathcal{N}t(\mu,\sigma; a, b)$ denotes the normal distribution $\mathcal{N}(\mu, \sigma)$ truncated at $a$ and $b$.}
\begin{tabular}{ |c|c|c|c|c| }
  \hline
   & \textbf{Parameter} & \textbf{Units} & \textbf{Definition} & \textbf{Prior} \\

  \hline
   & $M_{\text{tot}}$ & $M_{\odot}$ & Total mass $M_{\text{tot}} = M_{\text{halo}} + M_{\text{bar}}$ & Loguniform($\log(\mbar),\log(15.5)$) \\
   & $c$ &  -  & Halo concentration & Loguniform(0,2) \\
   & $\alpha$ & - & Inner slope of halo density & $\mathcal{U}(0,2.5)$\\
   & $\Upsilon_{\text{disc}}$ & $\msun / \text{L}_{\odot}$ & Disc  mass-to-light ratio & Lognormal (-0.72346, 0.24622) \\
   & $\Upsilon_{\text{bulge}}$ & $\msun / \text{L}_{\odot}$ & Bulge mass-to-light ratio & Lognormal (-0.38699, 0.24622) \\
   & $D$ & Mpc & Physical distance to galaxy &$\mathcal{N}\text{t}(\bar{D}, \delta D; 0, \text{None})$\\
   & $i$ & deg & Inclination ($0^\circ $ face on; $90^\circ$ edge on)  &  $\mathcal{N}\text{t}(\bar{i}, \delta i; 0, 180\deg)$   \\
  \hline
  & $M_{\HI{}}$ & $\msun$ & \HI{} mass (constrained by the flux profile) & $\mathcal{U}(0.5 \bar{M}, 1.5 \bar{M})$\\
  & $\sighi$ & kms$^{-1}$ & \HI{} velocity dispersion & $\mathcal{U}(0,100)$ \\
  & $\Sigma_0$ & $\msun \mathrm{pc}^{-2}$ & Maximum \HI{} surface density & $\mathcal{U}(0,25)$\\
  & $\overline{r}_b$ & - & Core size normalised by $r_{\HI}$ & $\mathcal{U}(0,1)$\\
  \hline
  & $\sigma_{\Psi}$  & $\textrm{Jy}$ & Additional Gaussian noise on measured flux & $\mathcal{U}(0,100)$  \\
  & $\sigma_{V}$  &  kms$^{-1}$ & Additional Gaussian noise on measured $V_{\mathrm{obs}}$ & $\mathcal{U}(0,20)$\\
  \hline
\end{tabular}
\end{table*}

\subsection{Assessing consistency}\label{sec:methods_consistency}

As a test of our flux profile model, we wish to assess whether the rotation curve data and the integrated flux profile data are consistent with a model in which they both have the same dark matter halo. We use the independence of the datasets to construct the total likelihood $\mathcal{L}_{\mathrm{tot}}= \mathcal{L}_{\Psi} \times \mathcal{L}_V$, which we use to constrain two models with both the rotation curve and flux data simultaneously:

\begin{enumerate}
    \item \emph{Joint halo model:} The flux profile and RC data are described by a shared set of halo and galaxy parameters.
    \item \emph{Separate halo model:} The flux profile and RC data are described by a shared set of galaxy parameters, but each of the two forward models have their own separate set of halo parameters, giving six halo parameters in total.
\end{enumerate}
We reject the separate halo model unless there is substantial evidence in its favour according to the Jeffreys scale ($\log(\mathcal{Z}_{\mathrm{separate}}/\mathcal{Z}_{\mathrm{joint}})$ > 0.5).

\subsection{A probabilistic model for $\Sighi$}\label{sec:probgas}

When fitting halo models to galaxies without resolved \HI{} data, it is necessary to model $\Sighi$, which will introduce an additional source of uncertainty. To this end we develop a probabilistic model based on the SPARC \HI{} surface density profiles, which can then be applied to larger samples. We do this by fitting a parameterised $\Sighi$ to the SPARC galaxies, and fitting a Gaussian Mixture Model to the distribution of parameters.

\citet{stevensOriginGalaxyHI2019}, in exploring the origin of the \HI{} mass--size relationship, tested three analytic profiles for the \HI{} surface density on a sample of 110 galaxies from the THINGS, LITTLE THINGS, LVHIS, and Bluedisk surveys, finding all three were a good fit to observations. We adopt their ``model 2'', which was found to be the best-fit to the most galaxies. Towards the centre the profile has a core with variable size, and at large radius it falls off exponentially with radius squared (a Gaussian),

\begin{equation}
\Sigma_{\rm H\,{\LARGE{\textsc i}}}(r) =
\left\{
\begin{array}{l r}
\Sigma_0, & r \leq r_b\\
\Sigma_0 \exp\left[-r_S^{-2} (r-r_b)^2 \right], & r > r_b
\end{array}
\right.\,,
\label{eq:model2}
\end{equation}
where $\Sigma_0$ is the central surface density, $r_b$ is the core size, and $r_S$ is the scale length of the exponential decline.  We define normalised lengths and surface densities

\begin{subequations}
\label{eq:bar}
\begin{equation}
\bar{r}_x \equiv r_x / r_{\rm H\,{\LARGE{\textsc i}}}\,,
\end{equation}
\begin{equation}
\bar{\Sigma}_x \equiv \Sigma_x / \left(1\,{\rm M}_{\odot}\,{\rm pc}^{-2}\right)\,,
\end{equation}
\end{subequations}
where $r_{\rm H\,{\LARGE{\textsc i}}}$ is the radius at which the gas drops to $\left(1\,{\rm M}_{\odot}\,{\rm pc}^{-2}\right)$.
The radius $r_{\rm H\,{\LARGE{\textsc i}}}$ can then be written as

\begin{subequations}
\label{eq:sizemass2}
\begin{equation}
r_{\rm H\,{\LARGE{\textsc i}}} = \sqrt{\frac{m_{\rm H\,{\LARGE{\textsc i}}}}{\pi\,\Sigma_0\,\left[\bar{r}_b^2 + \bar{r}_S\,(\bar{r}_S+\sqrt{\pi}\bar{r}_b) \right]}}\,,
\end{equation}
and
\begin{equation}
\bar{r}_S = \frac{1-\bar{r}_b}{\sqrt{\ln(\bar{\Sigma}_0) }}\,.
\label{eq:xip}
\end{equation}
\end{subequations}
We fit equation \ref{eq:model2} to the \HI{} surface density profiles of the SPARC galaxy, and find a similar distribution in model parameters to \citeauthor[][]{stevensOriginGalaxyHI2019} (their fig.~5). Following \citeauthor[][]{stevensOriginGalaxyHI2019} we assume $\Sigma_0$ and $\bar{r}_b$ are independent of other galaxy properties (i.e. self-similar density profiles). We check this is a reasonable assumption using SPARC, finding the strongest correlation is a weak (Pearson correlation coefficient of $0.26$) and not quite statistically significant ($p$-value $0.08$) correlation between $\Sigma_0$ and $M_{\mathrm{HI}}$. We make $\Sigma_0$ and $\bar{r}_b$ free parameters in the inference, with a prior given by the 2D distribution of ($\Sigma_0$,$\bar{r}_b$) for the SPARC galaxies, which together with the observed $M_{\mathrm{HI}}$ of each galaxy specifies the \HI{} surface density in full. To smooth and interpolate the parameter distribution we fit a Gaussian Mixture Model. We adopt the number of Gaussians that minimises the Bayesian Information Criterion (BIC; \citealt{schwarzEstimatingDimensionModel1978}), an approximation to the Bayesian evidence.

\subsection{Quantifying information content}

We wish to compare how much information on halo properties is obtained from using the full flux profile to constrain the inference versus the resolved RC or linewidth summary statistic. The information gain achieved in an experiment when transitioning from a prior to a posterior distribution can be quantified using the Kullback--Leibler (KL) divergence \citep[$\dkl$,][]{kullbackInformationSufficiency1951}. This measure is expressed in bits and is defined as:
\begin{equation}
D_\text{KL}(P \parallel \pi) = \int P(\theta) \log_2\left(\frac{P(\theta)}{\pi(\theta)}\right)\ \operatorname{d}\theta
\end{equation}
where $P(\theta)$ represents the posterior distribution and $\pi(\theta)$ the prior distribution. The KL divergence assesses the dissimilarity between these two distributions, effectively measuring the additional information provided by the posterior compared to the prior. This metric is particularly useful for evaluating the enhancement in constraint precision between different experiments \citep{buchnerIntuitionPhysicistsInformation2022}, as it takes into account the entire probability distributions rather than relying on summary statistics like credible intervals. However, this can make interpretation less straightforward. A simple illustration of the KL divergence is an experiment with a uniform prior that yields a uniform posterior with a hypervolume $k$ times smaller, resulting in a KL divergence of $\log_2(k)$.
For our KL divergence calculations, we employed kernel density estimation (KDE) to approximate the posterior probability distribution. Specifically, we utilised the fastKDE algorithm \citep{obrienReducingComputationalCost2014,obrienFastObjectiveMultidimensional2016}, which determines the optimal kernel and bandwidth based on established criteria  \citep{bernacchiaSelfconsistentMethodDensity2011}. To ensure the convergence of our KL divergence values, we verified that the results remained consistent across chains of varying lengths.
The KL divergence is a prior-dependent quantity. While our prior bounds are well-justified for mass and the lower bound of concentration, the upper bound of concentration is somewhat arbitrary. However, as our analysis focuses on relative KL divergences between different measurements, our conclusions remain robust to the specific choice of prior. \tys{motivate the bounds more maybe}

\subsection{Quantifying asymmetry}\label{sec:asymmetry}

Our model assumes axisymmetry, and therefore model misspecification may cause biased constraints in strongly asymmetric galaxies. To study this we use the asymmetry statistic developed in \citet{yasinConsistencyRotationCurves2024}, which we summarise here. We fit each flux profile using the ``generalised busy function'' \citep{westmeierBusyFunctionNew2014}, which is specifically designed to provide excellent fits to observed flux profiles

\begin{align}
    B(x) &= \frac{a}{4}
            \times (\mathrm{erf}[b_{1} \{ w + x - x_{\rm e} \} ] + 1) \nonumber \\
         &  \times (\mathrm{erf}[b_{2} \{ w - x + x_{\rm e} \} ] + 1)
            \times \left( c \, |x - x_{\rm p}|^{n} + 1 \right) \!,
    \label{eq:busyfunction}
\end{align}
where $a$, $b_1$, $b_2$, $x_e$, $x_p$, $c$, $w$, and $n$ are free parameters, and $\mathrm{erf}$ represents the error function. The function is able to describe both symmetric and asymmetric profiles, with symmetry imposed by setting $b_1 = b_2$ and $x_p = x_e$.




To quantify the degree of asymmetry, we estimate the evidence ratio for the full 8-parameter model and restricted 6-parameter symmetric version, $\mathcal{Z}_{\mathrm{asym}}/\mathcal{Z}_{\mathrm{sym}}$, using the BIC of each model. A larger $\mathcal{Z}_{\mathrm{asym}}/\mathcal{Z}_{\mathrm{sym}}$ value indicates a more pronounced asymmetry in the flux profile. Adopting a similar criterion to \citet{yasinConsistencyRotationCurves2024}, we consider overwhelming evidence for asymmetry if $\log(\mathcal{Z}_{\mathrm{asym}}/\mathcal{Z}_{\mathrm{sym}}) > 2.5$.

\section{Results}\label{sec:results}

\begin{figure*}
\includegraphics[width=0.8\textwidth,height=0.16\textheight]{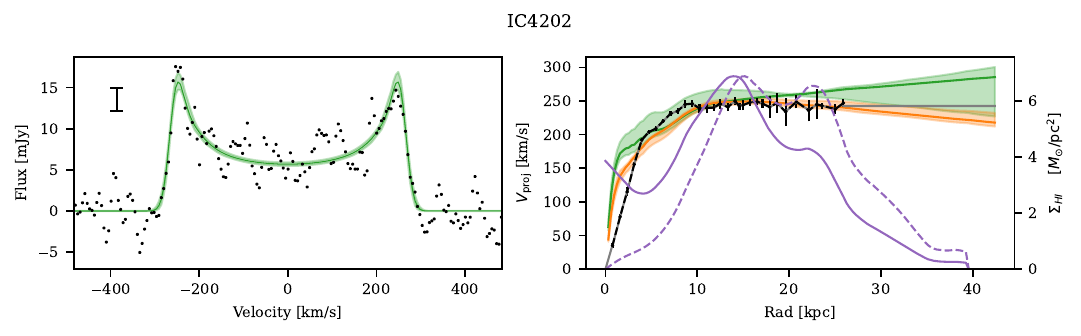}
\includegraphics[width=0.8\textwidth,height=0.17\textheight]{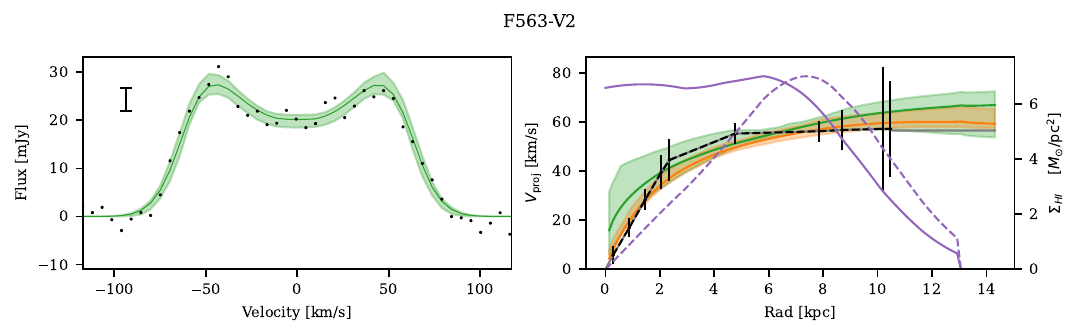}
\includegraphics[width=0.8\textwidth,height=0.17\textheight]{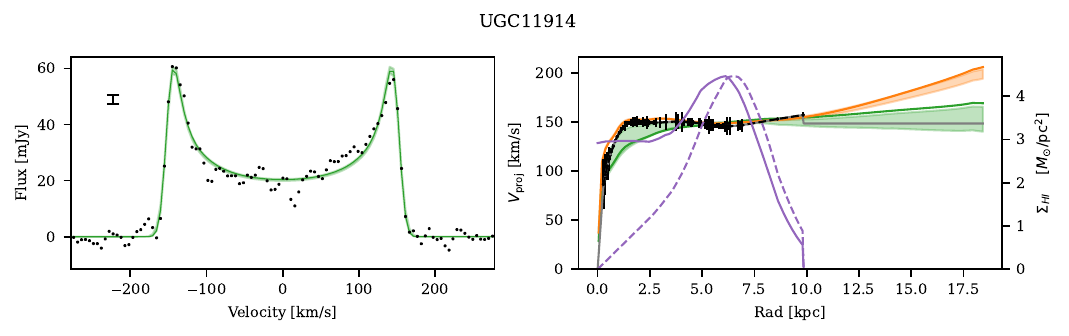}
\includegraphics[width=0.8\textwidth,height=0.17\textheight]{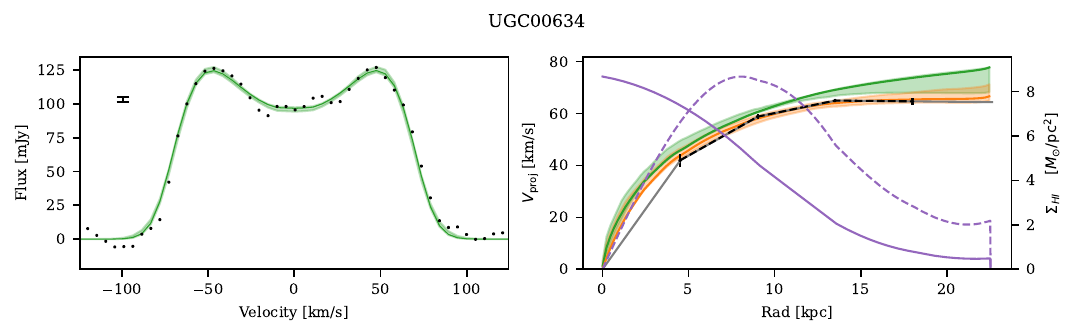}
\includegraphics[width=0.8\textwidth,height=0.17\textheight]{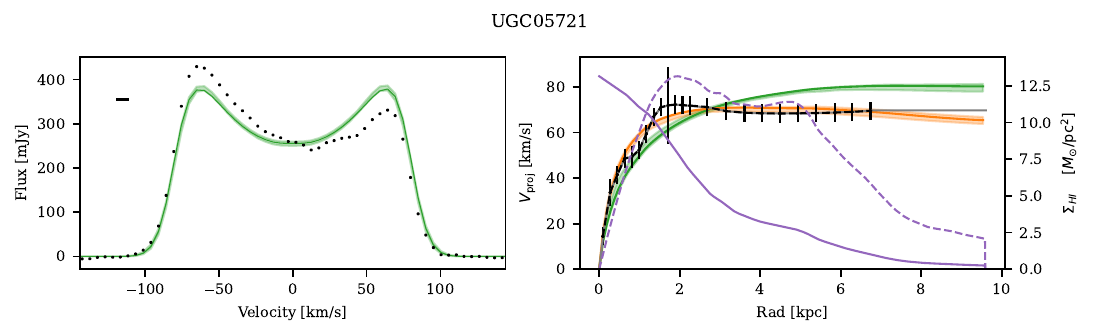}
\caption{\emph{Left panel}: The measured ALFALFA flux profile (black), with the RMS noise shown by the errorbar. Green shows the best fit model flux profile, with the shaded region the $1\sigma$ confidence interval. \\
\emph{Right panel:} The observed SPARC RC and uncertainty (black), with the best fit RC obtained from fitting to the RC shown in orange (with $1\sigma$ confidence region shaded). In green is shown the best fit RC obtained from constraining halo and galaxy parameters \emph{using the flux profile}, without any information from the observed RC. The solid purple line is the \HI{} surface density $\Sighi$, and the dashed purple line is $\propto r$$\Sighi$, its product with radius. The latter is proportional to the \HI{} flux at each radius, and hence shows which region of the RC the line profile is sensitive to.
\\
For the first 4 galaxies the RC obtained from the flux profile is in good agreement with the observed RC, showing the power of the integrated flux profile to constrain the shape of the RC, including its flatness at large radius corresponding to significant amounts of dark matter in the galaxy outskirts. UGC05721 on the other hand has a heavily asymmetric flux profile which the model is unable to match, resulting in an inferred circular velocity profile which is in disagreement with the observed RC, although the general shape is reproduced.
We show the remaining galaxies in the sample in Fig.~\ref{fig:app}.}
\label{fig:fit}
\end{figure*}

\begin{figure*}
\includegraphics[width=0.97\textwidth]{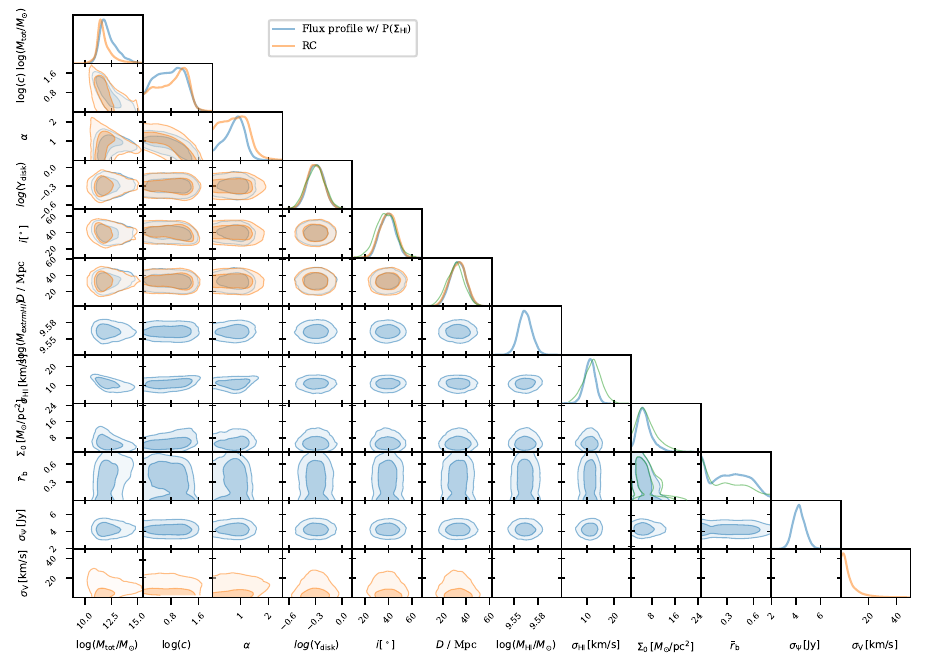}
\caption{\tys{MHI fix posterior} The full posterior on model parameters for galaxy UGC00634 for the inference constrained by the RC (orange) and integrated flux profile (blue), using the probabilistic $\Sighi$ model for the latter (see Section \ref{sec:probgas}). The model parameters are described in Table \ref{tab:parameters}. We show non-uniform priors in green, in 2D for $\Sigma_0$ and $\rb$ which have a correlated prior, and on the histogram only for the rest. Auxiliary galaxy parameters are generally well constrained, following their priors for the flux profile. There are degeneracies between halo parameters for both the RC and flux profile, with the RC providing improved constraints on the total mass $\mtot$ compared to the flux profile.}
\label{fig:posterior}
\end{figure*}

\subsection{Dark matter constraints}\label{sec:constraints}

The main output of this work is the constraints on the circular velocity profile and on dark matter halo parameters obtained from the inference with the spatially integrated \HI{} flux profile. We present the fits to the integrated flux profile and the projected circular velocity for some example galaxies in Fig. \ref{fig:fit}. The left column shows the flux profile data, with the best fit model and its $1\sigma$ confidence interval. The right panel shows the circular velocity profile calculated from the \emph{a posteriori} model parameters of this inference (\emph{green}). For comparison, we also show the best fit circular velocity and its uncertainty when using the SPARC RC to constrain the model instead (\emph{orange}). The solid purple line shows the \HI{} surface density profile $\Sighi(r)$, and the dashed line shows $r \Sighi(r)$, which is proportional to the amount of \HI{} flux emitted from a given radius. This therefore shows which radii the \HI{} flux profile is most sensitive to, which tends to peak roughly where the observed RC becomes flat.

For all galaxies the best fit flux profile is a good approximate visual match to the data. For the first four galaxies the rotation curve inferred from the flux profile is also a good match to the observed RC. As shown in \citet{yasinInformationHaloProperties2023}, for high mass galaxies which are baryon dominated towards their centre (\emph{maximal discs}), the same linewidth can be produced by both a galaxy with dark matter in the outskirts (leading to a flat RC), and a galaxy without any dark matter at all, because both have RCs that peak at approximately the same value. However by using the full flux profile this degeneracy is broken, as the full shape of the flux profile gives information on the flatness of the RC at large radii. In particular, a galaxy with a flat RC has a much greater ratio between the peak of the horns and the depth of the central trough compared to a galaxy lacking dark matter, which allows the RC to be well constrained.


UGC05721 is an example of a galaxy with a heavily asymmetric flux profile for which the model is unable to match the data. Whilst the inferred circular velocity still rises and flattens, the detailed shape and velocity of the flat part is in strong disagreement with the observational data.

Generally the constraints on the circular velocity from the RC inference are tighter than from the flux profile, as expected from spatially resolved data with independent datapoints at each radius, allowing the degeneracy between halo shape parameters to be broken. This is especially true at low radius, where the flux profile has little sensitivity, as shown by the dashed purple line. As expected the green band is narrowest approximately at the point where $r \Sighi$ peaks.

For UGC11914 we see that fit to the RC actually prefers a rising circular velocity profile out to large radius (likely driven by the outermost data point), whilst for the flux profile the RC is constrained to remain flat. This shows how RC measurements can commonly imply halo masses that are unexpected in the standard model, and hence, although we are treating the halo parameters obtained from the RC as a point of comparison, they are not necessarily the truth.

In Fig. \ref{fig:posterior} we show the full posterior for one fit constrained by the RC and one by the flux profile. Most parameters are well constrained, other than the halo parameters which display some strong degeneracies. The parameters describing the spatial distribution of the \HI{} gas follow the priors. The maximum \emph{a posteriori} Gaussian noise on the flux profile ($\sigma_\Psi$) is in the range of 0 to 5\% of the peak flux (although higher values are preferred for a couple of poorly fit galaxies). For the RC, $\sigma_V$ is lower, with a mean of \textasciitilde0.5\%.

\subsection{The consistency of RC and flux profile}\label{sec:results_consistency}

\begin{figure}
    \includegraphics[width=\columnwidth]{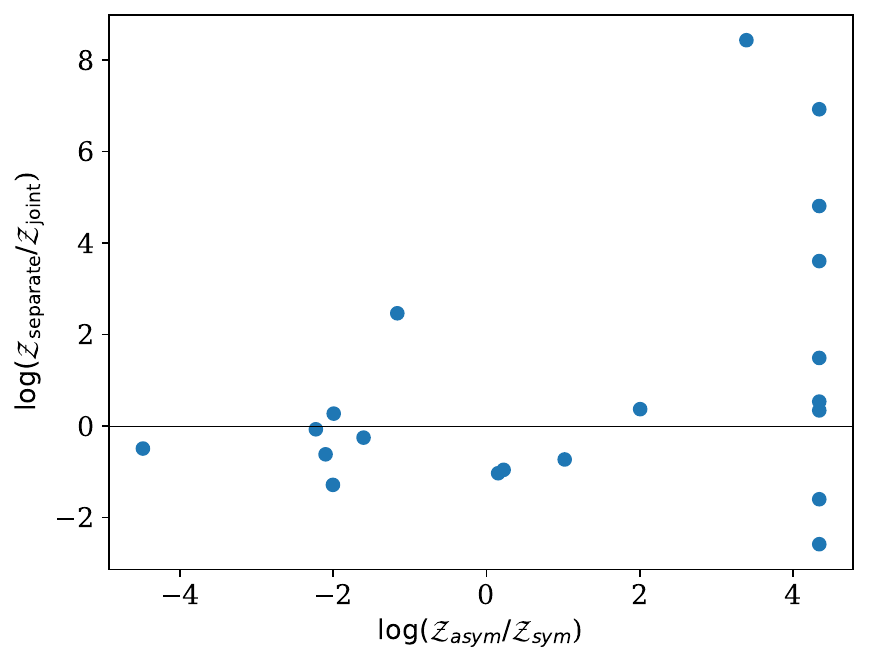}
    \caption{\tys{add guidelines for the eye}A plot showing the evidence for inconsistency between the flux profile and RC (quantified by the Bayes factor $\mathcal{Z}_{\textrm{separate}}/\mathcal{Z}_{\textrm{joint}}$, defined in Section \ref{sec:methods_consistency}) against the evidence for asymmetry (quantified by the $\zratio$ statistic, defined in Section \ref{sec:asymmetry}, and truncated to a maximum of 4.2). The separate model has a different set of halo parameters for the RC and flux profile, but the joint model has a single set. We reject the null hypothesis of consistency between the halo parameters inferred from flux profile and RC if $\log(\mathcal{Z}_{\textrm{separate}}/\mathcal{Z}_{\textrm{joint}}) > 0.5$, which corresponds to substantial evidence. All galaxies except one (F565-V2) show consistency as long as $\log(\zratio) \lesssim 2.5$, which corresponds to overwhelming evidence for asymmetry. F565-V2 is a galaxy for which either the RC or flux profile is likely badly measured (Section \ref{sec:results_consistency}).}
    \label{fig:agreement}
\end{figure}

\begin{figure*}
\includegraphics[width=0.6\textwidth,height=0.3\textheight]{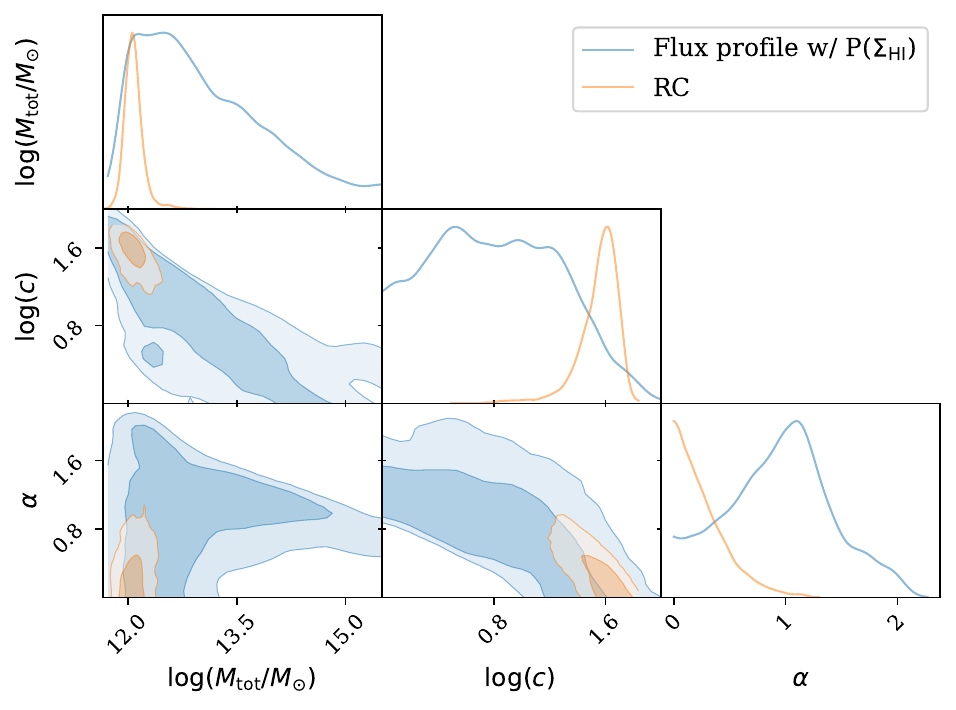}
\includegraphics[width=0.6\textwidth,height=0.3\textheight]{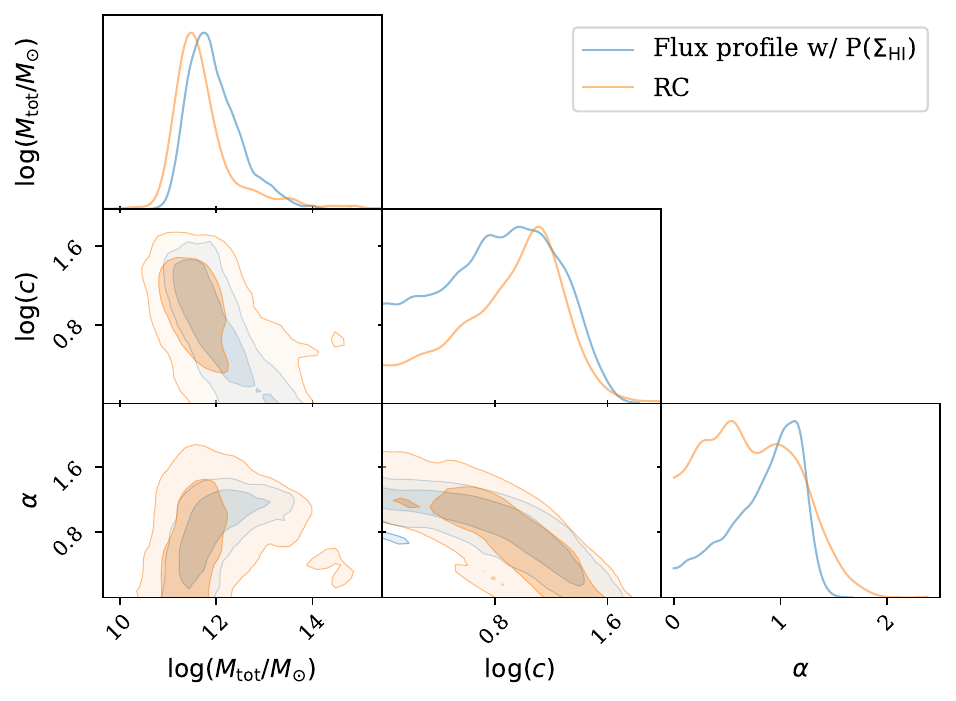}
\includegraphics[width=0.6\textwidth,height=0.3\textheight]{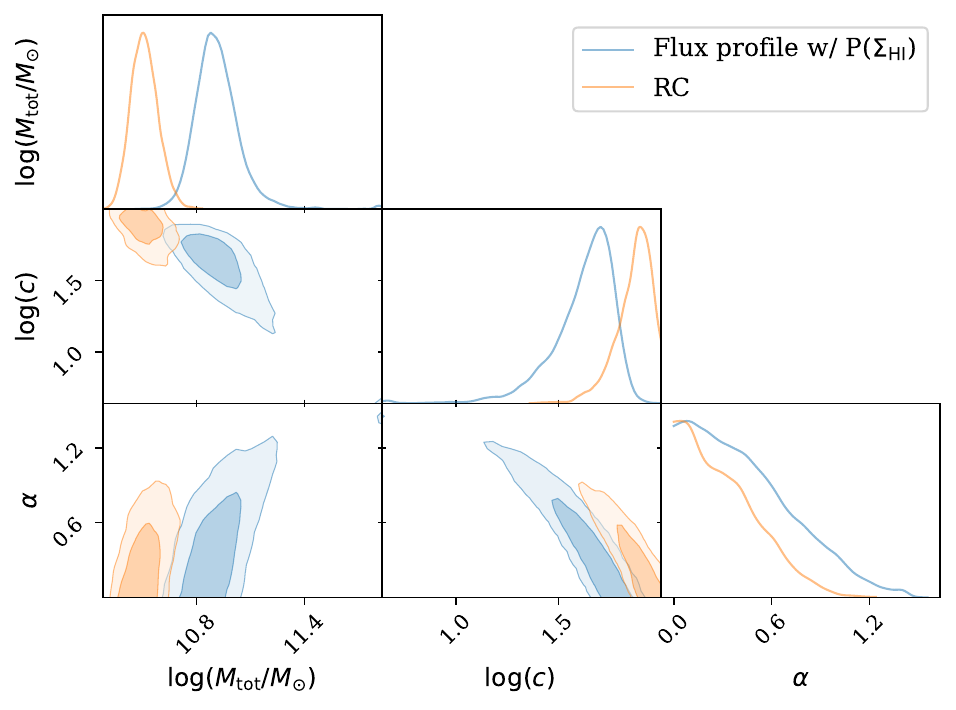}
\caption{The posteriors on halo properties for three example galaxies (\emph{from top}) IC4202, UGC00634 and UGC05721 for the inference constrained by the RC (\emph{orange}) and flux profile with probabilistic $\Sighi$ model (\emph{blue}). For IC4202 the flux profile provides interesting constraints on the 3-parameter space, but is not informative enough to break the degeneracies between total mass and shape parameters, unlike the RC. For UGC00634 the constraining power on halo properties is comparable between the flux profile and the full RC. For UGC05721, an example asymmetric galaxy for which the model cannot fit the flux profile well, the constraining power is comparable but the two posteriors are in disagreement, with the flux profile constraints likely biased.}
\label{fig:halopars}
\end{figure*}

To statistically assess the consistency between the halo parameters obtained from fitting to the RC and flux profile, we constrain two further models as described in Section \ref{sec:methods_consistency}. These are both joint fits to both the RC and flux profile simultaneously, but for the first \emph{joint} model, a single set of halo parameters is shared between both datasets, whereas for the \emph{separate} model, each dataset constrains a different set of halo parameters. Both datasets share a common set of galaxy parameters in both fits. The evidence for a separate halo over a joint halo is then the evidence ratio $\mathcal{Z}_{\mathrm{separate}}/\mathcal{Z}_{\mathrm{joint}}$.

 We display this statistic for the sample in Fig. \ref{fig:agreement}. We see that for all galaxies that are not statistically certain to have a very large asymmetry ($\log(\zratio) > 2.5$), $\mathcal{Z}_{\mathrm{separate}}/\mathcal{Z}_{\mathrm{joint}}$ is around 1 or less, showing a consistent RC is derivable from the flux profile alone. The one exception to this is F565-V2, which has a $\log(\zratio) \sim - 1.2$, but $\log(\mathcal{Z}_{\mathrm{separate}}/\mathcal{Z}_{\mathrm{joint}}) > 2$. We identified this galaxy as having clear inconsistency between the observed flux profile and RC \citep{yasinConsistencyRotationCurves2024}, likely due to either the RC or the flux profile being badly measured. We conclude as long as a galaxy is not strongly asymmetric ($\log(\zratio) \lesssim 2.5$), then the flux profile and RC halos are consistent, providing evidence that for this subsample of galaxies the constraints on the halo parameters from the flux profile are unbiased. We then repeat this exercise using the probabilistic model for $\Sighi$, rather than the measured SPARC value, finding the result is unchanged. This suggests that the probabilistic model does not introduce biases into the inferred halo parameters.

In Fig. \ref{fig:halopars} we show example posteriors on halo properties for the fiducial models (where either the flux profile \emph{or} the RC are separately used to constrain halo and galaxy parameters). As is typical for many galaxies in the sample, whilst the flux profile is still informative, the strong degeneracy between the two halo shape parameters and the total mass of the sample remains, showing for most galaxies in our sample the flux profile alone is not able to constrain the shape of the RC. However for UGC00634 the constraints on the halo parameters is in fact of similar strength to the RC, showing that using our assumptions a very well measured flux profile can provide comparable information. As the RC and flux profile are independent, the two sets of constraints can be combined (see \ref{sec:methods_consistency}) to give an improved joint constraint on halo properties. Finally we show UGC05721, a galaxy with an asymmetric flux profile for which the posteriors are visibly (although not strongly) in disagreement, and which our evidence based metric identifies as preferring inconsistent halo parameters for the RC and flux profile.

\subsection{The information content of the flux profile versus the RC}\label{sec:information}

\begin{figure}
    \includegraphics[width=\columnwidth]{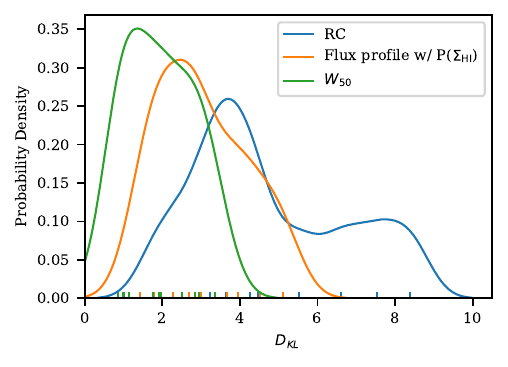}
    \caption{ \tys{Show joint constraints - have discussed in text only. dkl units} A plot showing the distribution of KL divergences $\dkl$ of the posterior of halo properties from the prior when constrained using the RC, the integrated flux profile (with probabilistic $\Sighi$ model, described in Section \ref{sec:probgas}) and the $W_{50}$ linewidth summary statistic. The ticks on the x-axis show the separate galaxies in the sample, while the curves are the KDE-smoothed probability distributions. Only galaxies matching our non-asymmetry criterion are included, which guarantees the RC and flux profile are consistent. On average the mean $\dkl$ increases by about 1.5 from the linewidth to flux profile, corresponding to a posterior about 3 times tighter posterior. The tightest constraints are still only obtainable from the RC however.}
    \label{fig:dkl}
\end{figure}

We now wish to assess the information profile for the population. In Fig. \ref{fig:dkl} we show the Kullback--Leibler divergence of the three dimensional ${\mtot, c, \alpha}$ posterior from the prior for the RC fit, the fit to the linewidth $W_{50}$ and the fit to the full flux profile using the probabilistic model for the \HI{} surface density, which we find does not significantly alter the constraints compared to the SPARC measured $\Sighi$. As expected, the RC has the largest amount of information. Going from $W_{50}$ to the full flux profile results in a shift in mean $\dkl$ of $\sim 1.5$, which corresponds to roughly a three times smaller posterior. This shows that a large amount of information is contained in the full flux profile relative to the linewidth, although the tightest constraints on halo parameters ($\dkl > 6$) are still only available from well measured RCs. We attempt to measure correlation coefficients between the $\dkl$ and galaxy/measurement properties. Although we find no statistically significant ($p < 0.05$) correlations due to the small sample size, the strongest non-significant correlation is with the signal-to-noise of the flux profile.

\section{Discussion}\label{sec:discussion}

\subsection{Overview}

We have presented a Bayesian model to constrain halo properties using the full shape of the spatially integrated \HI{} flux profile. We find, for galaxies without strongly asymmetries in their flux profiles (defined using the objective Bayesian criterion described in Section \ref{sec:asymmetry}), we are able to obtain constraints on gNFW halo parameters that are consistent with classic RC fitting. Specifically, we find that the full flux profile is able to constrain the flatness of the RC out to similar radii as RC studies for most galaxies (Fig. \ref{fig:fit}). This breaks the degeneracies in the constraints on halo properties from $W_{50}$ identified in \citet{yasinInformationHaloProperties2023}, which are especially prominent in baryon-dominated high mass spiral galaxies.

Initially, we used the resolved \HI{} surface density measured by SPARC in the forward model of the flux profile. However the overall goal of this project is to provide a method to constrain dynamics using \HI{} even in regimes where no resolved observations are available, such as at high redshift or in low angular size galaxies. Therefore we developed an empirical probabilistic model for the \HI{} surface density, which generalises \HI{} scaling relationships such as the mass--size relationship, to produce a full probabilistic model for the distribution of \HI{} given the mass. We find that replacing the measured SPARC $\Sighi$ for each galaxy with the probabilistic model in the inference does not lead to biased constraints and does not degrade the information content significantly, paving the way for application to datasets without resolved \HI{} observations. We find the flux profile inference with probabilistic $\Sighi$ produces posteriors around three times tighter than when the linewidth summary statistic $W_{50}$ is used as the observable.

The potential of mass modelling using integrated flux profiles has been only briefly explored in the literature. \citet{2021arXiv210504570P} applied a similar flux profile model to extract halo parameters for a pair nearby galaxies, using scaling relation-based models for the stellar and gas distribution and a simple chi-squared minimisation inference. They found halo mass and concentration parameters that they deemed plausible. We have conducted a rigorous study using state-of-the-art mass models and a full Bayesian inference over model parameters, along with a study of consistency with resolved rotation curves. Further we have introduced a probabilistic $\Sighi$ model that better captures the diversity of \HI{} distributions observed in real galaxies.

\subsection{Limitations}

We have assessed the accuracy of our integrated model by studying the consistency between the halo parameters inferred from RCs and flux profiles of a small sample of galaxies for which both measurements are available. The validity of our assessment therefore relies on both types of data being accurate. \HI{} astronomy is notoriously challenging, and it is possible that systematics are present in one or both datasets.

We obtained RCs from the SPARC database, the current state-of-the-art compilation of \HI{} RCs and mass models. However, it is still a compilation of inhomogenously measured and analysed RCs from literature. With future large surveys obtaining large homogenous samples of RCs \citep{degWALLABYPilotSurvey2022, maddoxMIGHTEEHIHIEmission2021}, it is likely that future datasets will be both larger and more reliable. We utilised flux profiles from the ALFALFA survey. The main limitation here is that the SPARC galaxies are nearby and have large angular size on the sky compared to the majority of the ALFALFA sample. Reconstructing the entire flux profile from galaxies with angular size larger than the beam width is notoriously difficult. To address this, we used galaxies from the ALFALFA reanalysis pipeline of \citet{hoffmanTotalALFALFANeutral2019} that is designed to best reconstruct the flux profile from these extended galaxies. However, \citeauthor{hoffmanTotalALFALFANeutral2019} acknowledge that the upgraded pipeline is likely still not able to reconstruct flux perfectly.

In the future it would be optimal to use galaxies with well measured RCs using next-generation interferometers that will be able to measure RCs at cosmological distances. Pan et. al (in prep) test the flux profiles obtained from MIGHTEE against FAST, finding on average that FAST galaxies have $7\pm 2 \%$ higher \HI{} mass. This suggests pipelines for current and future instrument are also better at reconstructing the flux profile even when the galaxy is larger than the beam size.


In the future, repeating this experiment with larger datasets will allow a more detailed study of both the consistency of RCs and flux profiles. For example, it is likely that at fixed profile asymmetry, different flux profile phenomenology (such as horned vs. non-horned profiles) will be more or less susceptible to systematic effects or failure in model assumptions. This could allow the construction of more elaborate criteria than a simple asymmetry cut for deciding which galaxies return consistent constraints under axisymmetric modelling of the flux profile.

\subsection{Model improvements}

An important future direction will be to modify the model to allow axisymmetry. This could include allowing an inclination warp, or angular variations in the \HI{} surface density. To construct parameterised asymmetric models, it will be useful to look at the 2D \HI{} surface density distribution in real galaxies, as well as the distribution of inclinations across the disc. \cite{pengParameterizedAsymmetricNeutral2023} constructed a simple physical model for asymmetric linewidths, by assuming the projected surface density varied with constant angular gradient across the sky. In the context of our approach, an asymmetric model for the intrinsic \HI{} distribution would have to be developed, with its parametrisation and orientation constrained as part of the inference. The kinematic model of \citet{pengParameterizedAsymmetricNeutral2023} is relatively simple, assuming the RC is deconstructed into a rising part giving a Gaussian profile, and a flat part giving the double horned shape. In our model we allow the detailed RC shapes obtained by varying the distribution of baryons and dark matter.

It should also be possible to use optical data on the 2D distribution of stars/gas to inform the distribution of \HI{} in galaxies. This would be similar to the method of \citet{scholteColdGasMass2023}, who use optical emission lines and photoionization models to reconstruct the cold gas surface density.

\subsection{Application}

The framework we have developed here can now be applied to samples of galaxies for which only spatially unresolved \HI{} observations are available. A particularly important use case will be extending the redshift range in which halo properties can be constrained with \HI{}. The FAST Ultra-Deep Survey (FUDS) recently measured flux profiles of six galaxies at $z>0.38$ \citep{xiMostDistantHI2024}. Although typical galaxies at cosmological distances will have lower flux profile SNR than many of the galaxies we have studied, future surveys such as MIGHTEE \citep{maddoxMIGHTEEHIHIEmission2021, ponomarevaMIGHTEEHBaryonicTully2021} will dramatically improve sensitivity and bring better control of systematic noise. Programmes are also underway to further extend the maximum distance \HI{} is detected to using gravitationally lensed sources \citep{deaneGravitationallyLensedHI2017,blecherNeutralHydrogenLensing2024}.

Due to the lack of sensitivity to the flux profile for the inner parts of the RC, we found the constraints on the halo shape parameters $\alpha$ and $c$ are weak, and sometimes biased in galaxies with asymmetric profiles. A powerful use of our method would be to combine integrated \HI{} with optical IFU data from surveys such as MaNGA \citep{bundyOverviewSDSSIVMaNGA2015} and WEAVE \citep{jinWidefieldMultiplexedSpectroscopic2024}, that can provide constraints on the inner part of the RC. Furthermore, the stellar kinematics can also provide a constraint on the galaxy inclination, another property that is typically measured from resolved \HI{} in current analyses. This is the subject of an upcoming study. Alternatively, high resolution data from instruments such as Euclid \citep{trojaEuclidNutshell2022} can be used to constrain the inclination from the optical disc.

Many studies are underway using IFU data to constrain dynamics at around redshift 1 \citep[e.g.][]{stottKMOSRedshiftOne2016}, using tracers that are limited to the inner regions of galaxies. Our method can also be adapted to model the integrated flux profiles of emission lines from other tracers of rotation such as molecular gas, that can be observed at very high-redshift using instruments such as ALMA.


A caveat to applying our method to new datasets is that the probabilistic $\Sighi$ model is derived from SPARC RCs, and it is possible a different \HI{} sample would have somewhat different \HI{} properties. However scaling relationships such as the \HI{} mass--size relation have been studied in various samples, suggesting this should not be a significant problem. Despite the \citeauthor{stevensOriginGalaxyHI2019} model being relatively simple (and not able to capture the full complexity of the \HI{} surface density), our probabilistic model is successful because the flux profile is relatively insensitive to the inner regions of the galaxy (see Section \ref{sec:constraints}). \tys{Discuss photoionization of \HI{} at higher redshift}

Given the relative simplicity of measuring flux profiles compared to RCs \citep[e.g.][]{roperDiversityRotationCurves2023}, assessing the consistency of constraints obtained from well-measured flux profiles with their RC counterparts could be a good way to identify potentially unreliable RCs for particular galaxies of interest. As the two sets of observations are fully independent, they can also be combined to obtain improved constraints, as done in our \emph{joint inference} (Section \ref{sec:methods_consistency}).

The methods developed in our mass modelling approach could also be repurposed for forward modelling of galaxy populations. Historically forward modelling has been carried out using summary statistics of scaling relations \citep[e.g.][]{desmondTullyFisherMasssizeRelations2015}. A model with galaxy-by-galaxy predictions for the full flux profile would likely be highly informative at higher redshift. Whilst our study has primarily targeted dark matter halos, it would also be possible to constrain other dynamical models such as modified gravity using this method for which there is interest in redshift evolution \citep[e.g.][]{hossenfelderRedshiftDependenceRadialAcceleration2018,naikConstraintsChameleonGravity2019, braxTestingScreenedModified2021, landimTestingScreenedModified2024}.

Whilst we have only applied the inference pipeline to a small galaxy sample, it is computationally cheap enough to apply to large samples such as ALFALFA ($\sim$30,000 galaxies). Each likelihood evaluation takes $\sim 10^{-4}$s, resulting in around one minute to complete the inference per galaxy.

\section{Conclusions}\label{sec:conclusions}

We introduce and validate a framework to constrain dark matter halo properties using the full shape of the spatially integrated \HI{} flux profile, as an alternative to RCs which are expensive to obtain and completely inaccessible in some regimes (particularly high redshift and low mass). Our Bayesian method consists of a full forward model for the spatially integrated \HI{} flux profile from parameters describing the halo and baryonic mass distributions, via a model circular velocity curve and \HI{} surface density distribution. We obtain halo constraints by applying the model to a sample of galaxies with both ALFALFA integrated flux profiles and SPARC resolved RCs, and test their relative information content and consistency.

\begin{itemize}
    \item For galaxies without significant asymmetries in their flux profiles, the inferred dark matter halo parameters from the flux profiles are consistent with those obtained from RCs. This validates the use of integrated flux profiles as a reliable tool for mass modelling in such cases.
    \item The full flux profile provides significantly more information on halo properties than equivalent use of the linewidth summary statistic ($W_{50}$). On average, using the full profile tightens the constraints on halo parameters by approximately three times compared to $W_{50}$. The RC still offers the most detailed information (particularly for constraining the inner regions of halos), although this is rivalled by the full flux profile in some cases.
    \item We introduce a probabilistic model for the \HI{} surface density profile, which can be applied to galaxies without resolved \HI{} observations. This model does not significantly degrade the information content of the flux profile or introduce biases into the inferred halo parameters, making it suitable for application in various contexts, including high-redshift studies.
\end{itemize}

The method we develop holds promise for extending dynamical studies of galaxies into regimes that are currently impossible to explore with traditional RC-based approaches due to limitations in sensitivity and resolution. In particular, as next-generation facilities such as the SKA come online, our framework can be applied to constrain the dark matter content of galaxies across cosmic time and to conduct statistical tests of the faint end of galaxy formation. We also show that \HI{} flux profiles contain significant \emph{complementary} information to RCs and other dynamical probes. In upcoming work we explore the potential of combining flux profiles with IFU stellar kinematics maximise leverage on galaxies' dark matter distributions.


\section*{Acknowledgements}

We thank Matt Jarvis and Federico Lelli for useful inputs and discussion. We also thank Federico Lelli for providing the SPARC \HI{} surface density data.

TY acknowledges support from a UKRI Frontiers Research Grant [EP/X026639/1],
which was selected by the ERC. HD is supported by a Royal Society University Research Fellowship (grant no. 211046). This project has received funding from the European Research Council (ERC) under the European Union's Horizon 2020 research and innovation programme (grant agreement No 693024).

For the purpose of open access, the authors have applied a Creative Commons Attribution (CC BY) licence to any Author Accepted Manuscript version arising.

\section*{Data Availability}

Data from the ALFALFA survey is located at \url{http://egg.astro.cornell.edu/alfalfa/data/index.php}. The SPARC database is located at \url{http://astroweb.cwru.edu/SPARC/}. Other data will be made available on reasonable request to the corresponding author.



\bibliographystyle{mnras}
\bibliography{new} 




\appendix

\section{Flux profile constraints}\label{sec:appendix_constraints}

A plot containing the remaining flux profile fits and RC constraints not shown in Fig. \ref{fig:fit}.

\begin{figure*}\label{fig:app}
\includegraphics[width=0.4\textwidth,height=0.12\textheight]{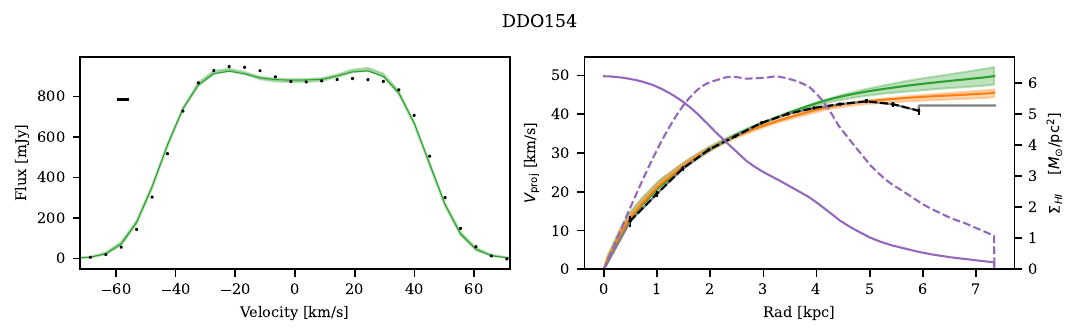}
\includegraphics[width=0.4\textwidth,height=0.12\textheight]{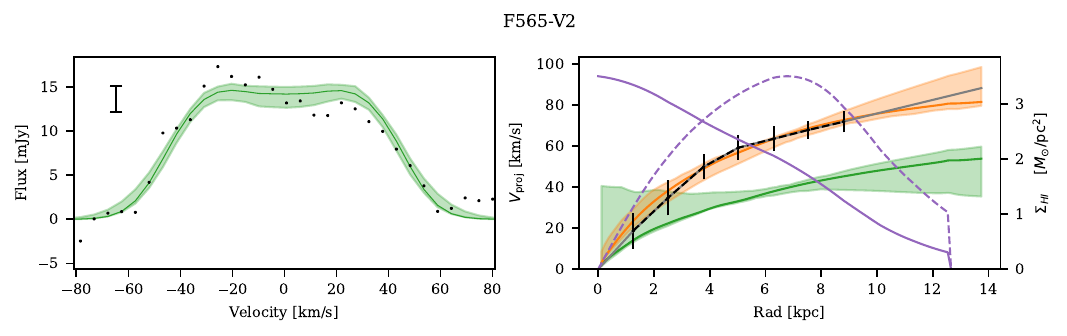}
\includegraphics[width=0.4\textwidth,height=0.12\textheight]{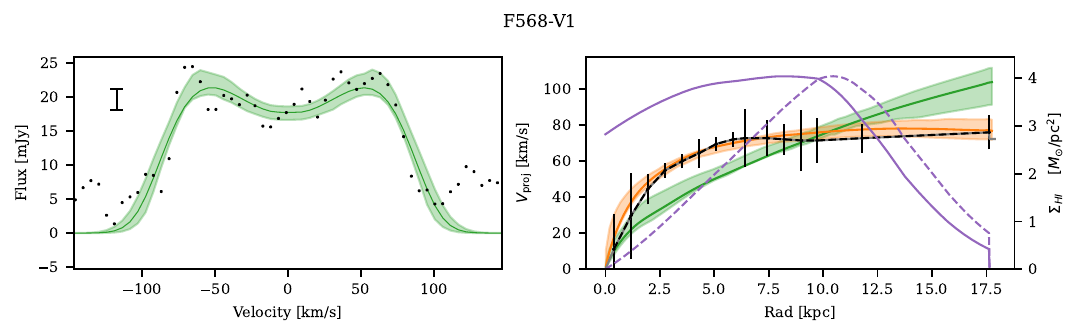}
\includegraphics[width=0.4\textwidth,height=0.12\textheight]{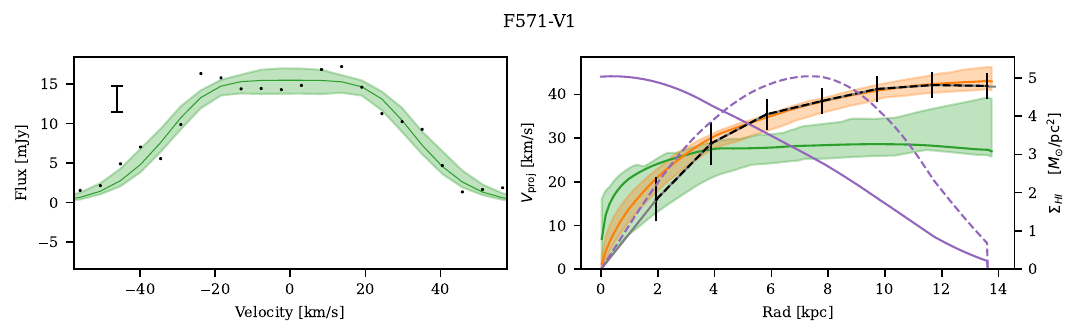}
\includegraphics[width=0.4\textwidth,height=0.12\textheight]{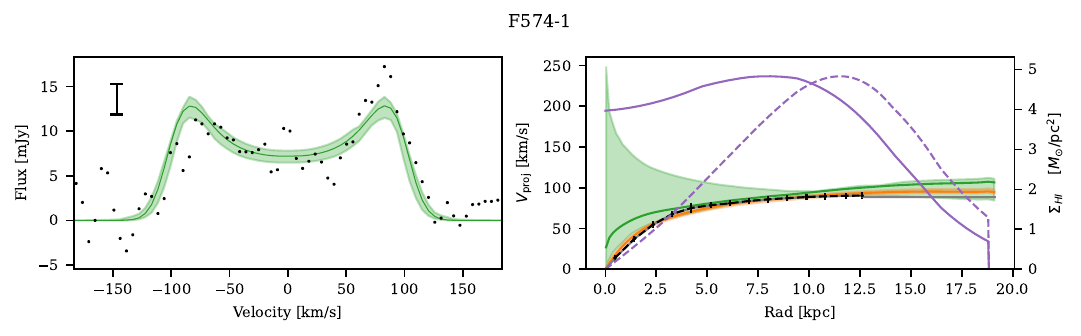}
\includegraphics[width=0.4\textwidth,height=0.12\textheight]{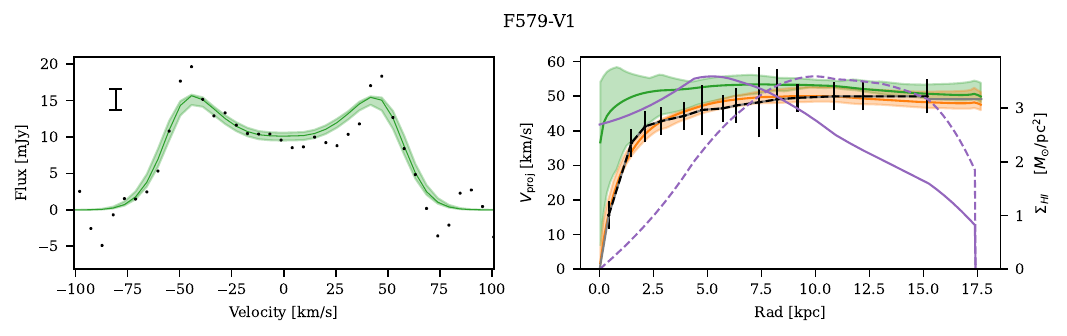}
\includegraphics[width=0.4\textwidth,height=0.12\textheight]{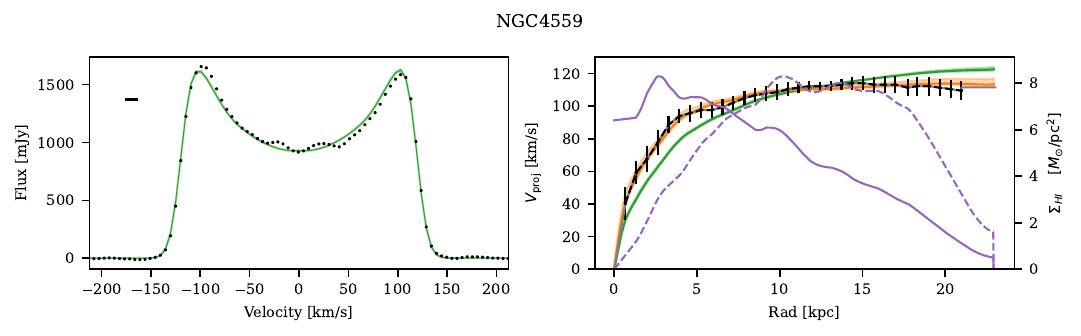}
\includegraphics[width=0.4\textwidth,height=0.12\textheight]{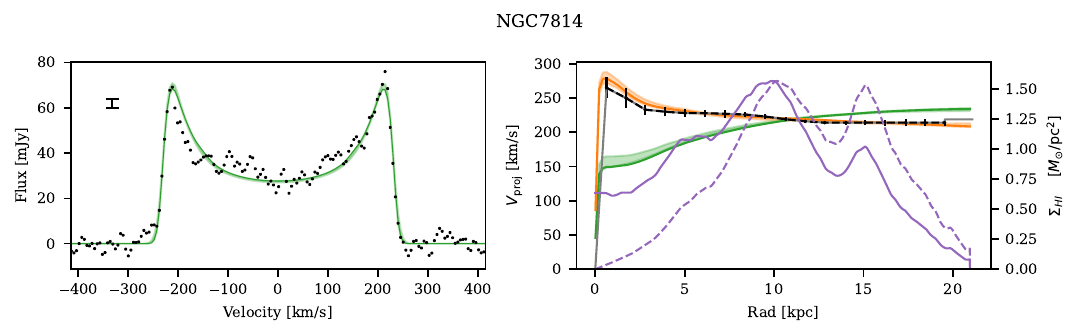}
\includegraphics[width=0.4\textwidth,height=0.12\textheight]{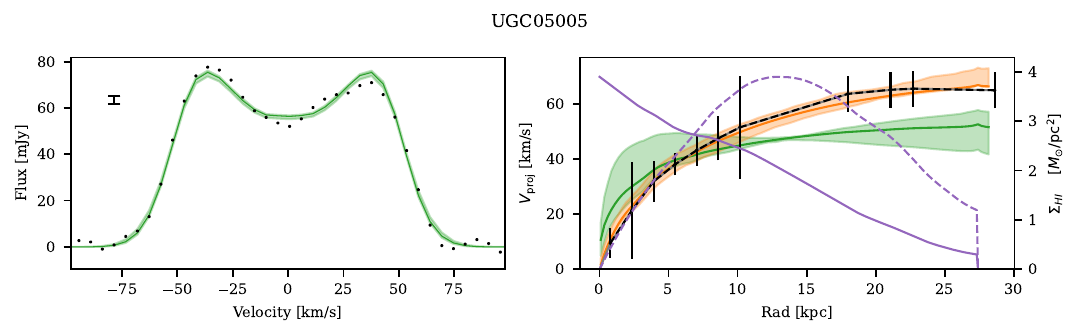}
\includegraphics[width=0.4\textwidth,height=0.12\textheight]{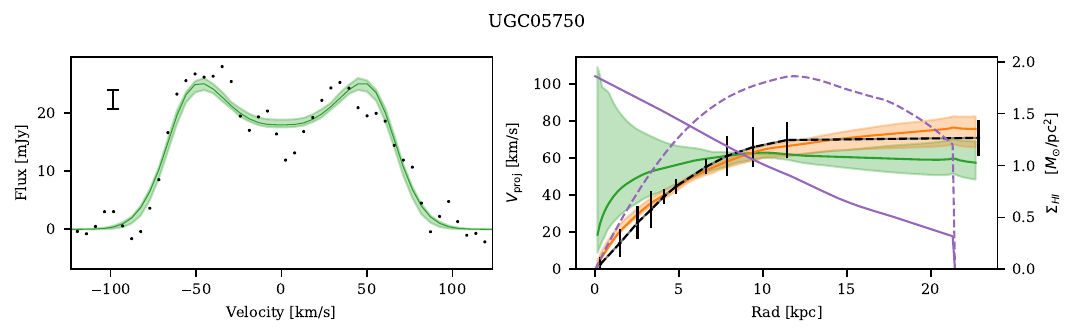}
\includegraphics[width=0.4\textwidth,height=0.12\textheight]{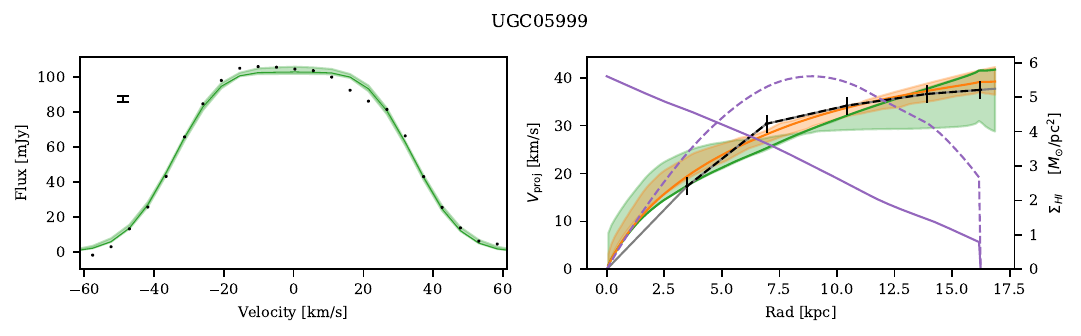}
\includegraphics[width=0.4\textwidth,height=0.12\textheight]{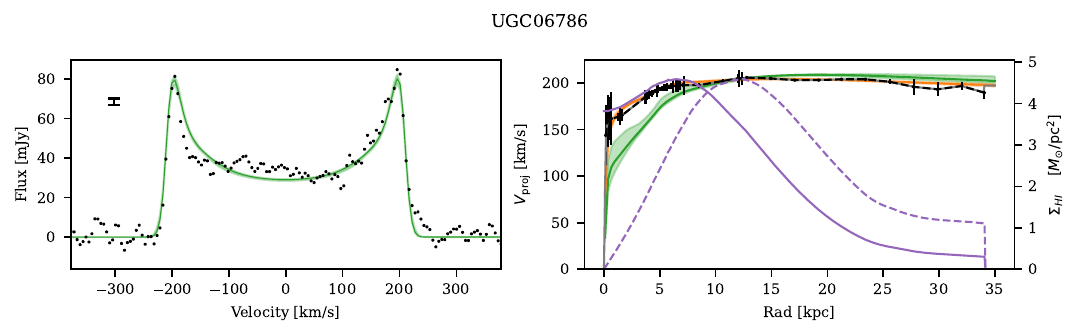}
\includegraphics[width=0.4\textwidth,height=0.12\textheight]{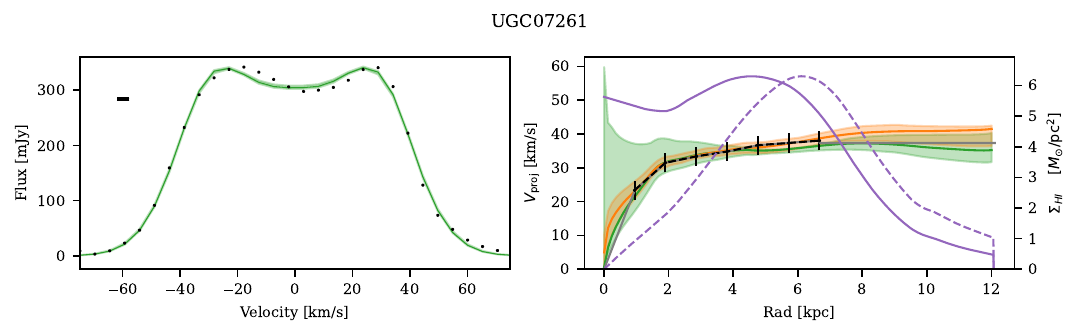}
\includegraphics[width=0.4\textwidth,height=0.12\textheight]{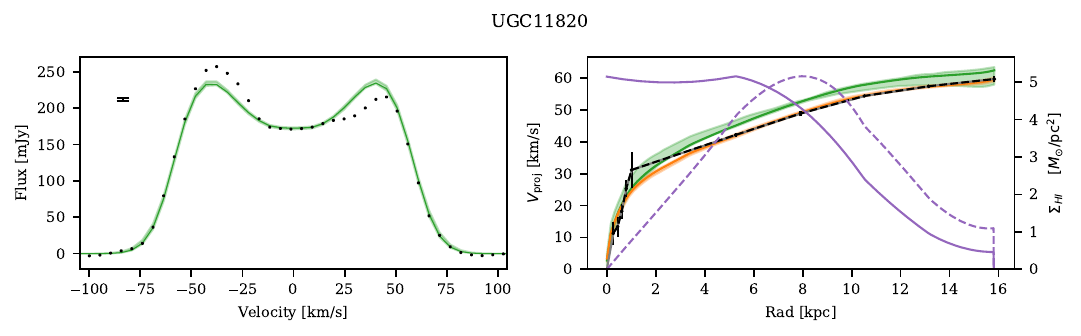}
\includegraphics[width=0.4\textwidth,height=0.12\textheight]{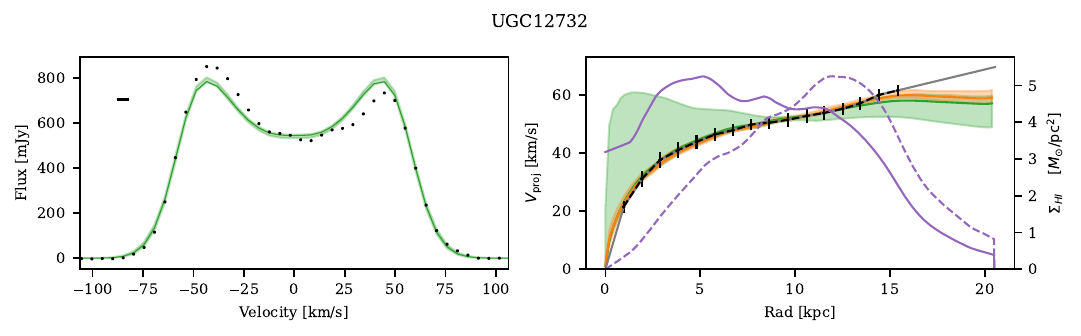}
\caption{As Fig.~\ref{fig:fit}, but for the other galaxies not shown there.}
\end{figure*}




\bsp	
\label{lastpage}
\end{document}